# MINING TRANSACTIONAL DATA TO PRODUCE EXTENDED ASSOCIATION RULES USING COLLABORATIVE APRIORI, FSA-RED AND M5P PREDICTIVE ALGORITHM AS A BASIS OF BUSINESS ACTIONS

Feri Sulianta[1st]
Department of Informatics Engineering, University of Widyatama, Bandung, Indonesia
feri.sulianta@widyatama.ac.id

Eka Angga Laksana[2nd]
Department of Informatics Engineering, University of Widyatama, Bandung, Indonesia
eka.angga@widyatama.ac.id

Thee Houw Liong[3rd] Departement of Physic, Bandung Institute of Technology, Bandung, Indonesia the@fi.itb.ac.id

*Abstract* − **There are large amounts of transactional data which showed consumer shopping cart at a store that sells more than 150 types of products. In this case, the company is utilizing these data in making business action. In previous studies, the data that has a lot of attributes and record data reduction algorithms handled by the FSA Red (Feature Selection for Association Rules) are then mined using Apriori algorithm. The resulting association rules have high levels of accuracy and excellent test results, which rely more than 90%.**
**In this study, the association rules generated in previous research will be updated by using prediction algorithms M5P, so that the reliability of association rules can be updated for the next day forward. Furthermore, some data mining technique such as: clustering and time series pattern will be implemented to examine the truth and to extend the validity of association rules which were built. It can be concluded that the association rules were established after will generate strong association rules with confidence equal or higher than 70% and the truth of the rules can be seen from the time series pattern on each group of goods which are then used as the basis of business actions.**

*Keyword-* **Association Rules, Apriori, Confidence, Clustering, Data Reduction, FSA-Red Algorithm, M5P, Time Series Patterns, Support**

## I. INTRODUCTION

The distinctive company who has large volumes of data is stored in the transaction processing system, which archived customer transaction histories related to the variations of products. The company wants to take advantage of data using data mining techniques in getting knowledge of transactional data as a basis to make business actions. Transactional data was held in the form of consumer shopping receipts archive of the range of products offered by the company.

According to the customers' choice regarding the product, when the consumers buy food A, they also buy food B, but on contrary they do not buy X items product if they buy items product Y or items product Z and vice versa. The Company wants to build business actions to address the customers' taste patterns.

The goals can be achieved by selling products that meet the tastes and consumers' behaviors based on the way they select the food products. The ability to read consumers' tastes is obtained using data mining techniques with method called Market Basket Analysis. This method shows people's decisions to buy products. However problems arise with the large number of data which will be used. There are 163 attributes with more than 90,000 records in 5 months. These will continue to expand along with the amount of data to be analyzed. Data preprocessing technique needs to determine specifically to handle large volumes of data to find the patterns of the food product consumers' tastes based on the sales transaction data in mining process.

The result of association rules after the mining process will be updated in the future with some span of prediction techniques, so that the association rules will remain usable and reliable for the long term. Furthermore, the mapping data into time series pattern on each group of goods and clustering analysis was conducted to see the truth and correlation of association rules.

## II. METHODOLOGY

The research design of this study includes data preprocessing, the specific algorithm selected, build the model and ended with evaluate the model to get justification regarding the procedure for mining the data, clustering technique and time series analysis. Each step will be described below:

*A. Data Preprocessing*

Data reduction process has been done on high volumes of transactional data to result the association rules which use as a basis of business actions. The complete stages from data set until business action consideration described in the block diagram below:

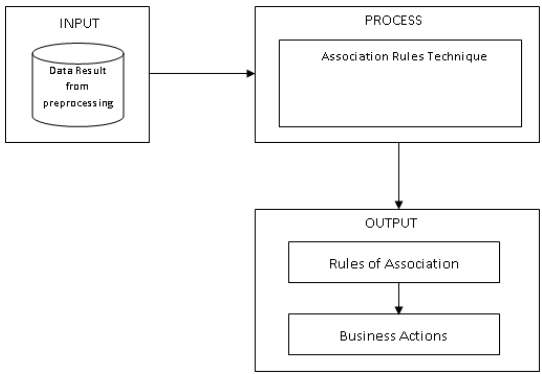

Fig. 1. Global block diagram of transactional high volumes data using apriori algorithm to produce association rules for business actions

Global block diagram in fig.1. shows component input is the result of data processing in a suitable form that is ready to do the mining. In the process stage, the data will be mined using apriori algorithm. As a result, the output stage will perform list of association rules which can be consider as a basis of a business actions.

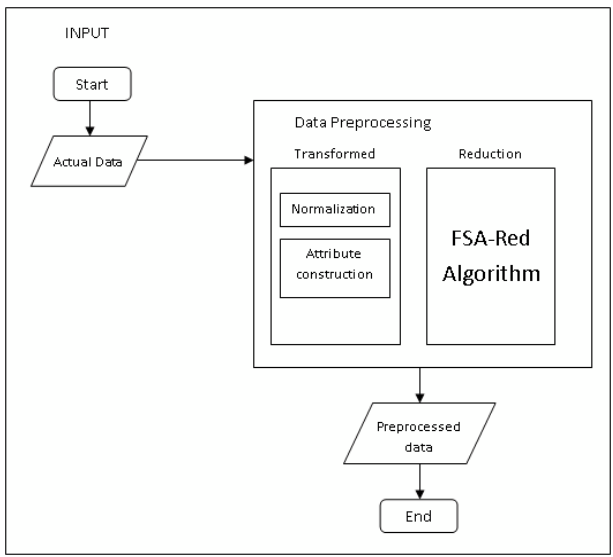

Fig. 2. Phase of input system

The actual data will be transformed using normalization and attribute construction. In the transformation process, value of attribute which is not 0 will be converted to 1 while the value of attribute 0 unchanged. It is intended to represent the presence and absence of product purchase. Thereafter representing 163 product attributes will be constructed then each dataset will clearly show all products.

*B. FSA-Red Algorithm*

Data that has been transformed using the algorithm will be reduced using FSA-Red algorithm. Attributes that will be analyzed will be the basis of data reduction. FSA- Red-algorithm will detect relationships between attributes in an instance and only instance that has no correlation with the attributes to be analyzed have to be eliminated, followed by eliminating attributes that do not want to be analyzed which named feature selection. The result of this stage is preprocessed data, which ready to be used for mining.

*C. Generating Association Rules*

Association rules will be generated using Apriori Algorithm collaborated with reduction procedure. In this stage, the association rules that are tested achieve confidence of> = 70% of the overall data. If confidence on association rules which achieve <70% then the association rules will be categorized into association rules are not validated and will be eliminated from the group confidence association rules that have> = 70% in the next stage.

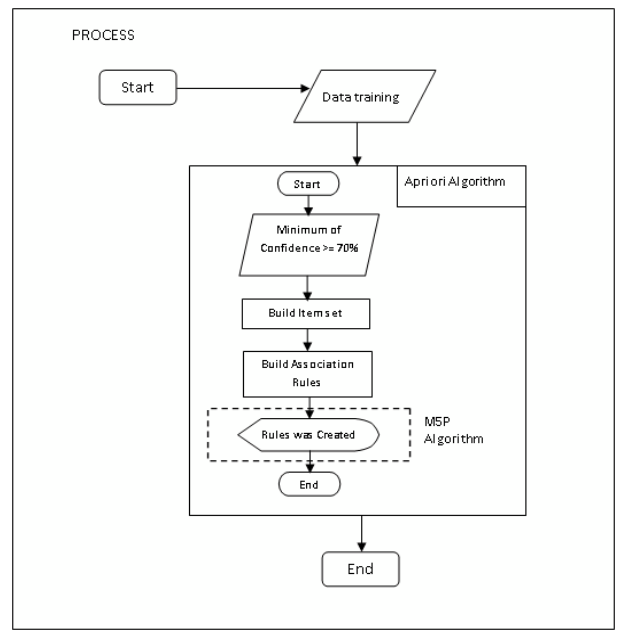

Fig. 3. Phase of apriori algorithm to build association rules

Preprocessed data is considered as data training. In this case, apriori algorithm will read and learn the repeated data pattern which referred to as frequent item set. Then, Association rules will be created based on the determined confidence equal or higher than 70%.

*D. Extended Association Rules*

To extend the reliability of the association rules that was been produced, prediction method using MD-5Algorithm will be implemented. M5P algorithm will analyze and make the value of output by values collection of instance and feature vectors, as well as the desired time span limitation prediction calculations.

Basically, this algorithm will make and implement reconstruction decision trees with linear regression function, for each node is formed. There are three major steps to do that:

1. Decision-tree induction intended to build logic tree, the branches will continue to be formed, and the execution process at this stage will stop if the instance is no longer found new variations.
2. The tree which already made the first step will be cut, with regression techniques.
3. The third step, which is to avoid fluctuating conditions and avoid inconsistencies, would smoothing technique will reconstruct the logic tree form, includes branches and leaves. This is done by combining the reconstruction process with a linear model that will produce a predicted value

*E. Extended Association Rules Pattern Analysis*

Association rules generated will be analyzed using the next month data and time series pattern's on group of goods. K-Means algorithm is used to construct the four groups of clustered data. Based on clustered, the value of attribute on each good will be translated on time series pattern.

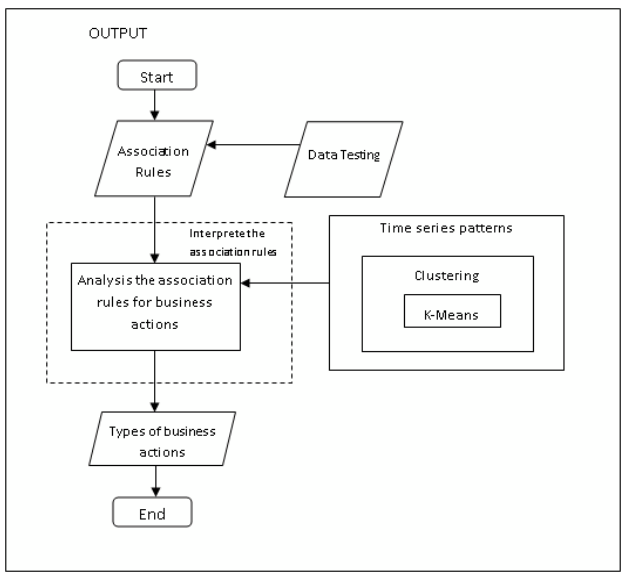

Fig.4 Phase of output system

Thus the business actions are made based on association rules regarding the characteristics of clustered time series pattern that can be implemented on each group of goods.

## III. RESULTS

Time-Series Analysis is implemented to assess the sales history of a product. Transactions every customer for the two best products showed number of these products per transaction. The result did not reflect a clear chart patterns, because on the other hand there is a product that was not purchased that day and the number of products purchased were not more than two items of products in general. According to the condition, the product will be group with aggregation method per day transaction.

Thus the treatment for grouping performed by classifying the number of selling per day. Feature selection is used to choose the most high rate products is calculated by the number of sales per day.

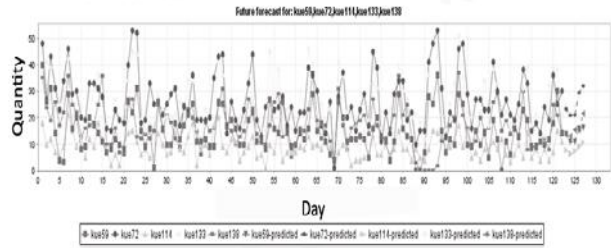

Fig.5 Time series analysis on the best-selling items product daily

Graph analysis shown the extended association rules in certain cluster, which is a group of products with the highest sales level. In this case the process of clustering using K-Mean algorithm. The results of time series of data on this cluster shows the correctness of the implementation of the rules of the association purchased revealed if X then Y is not purchased.

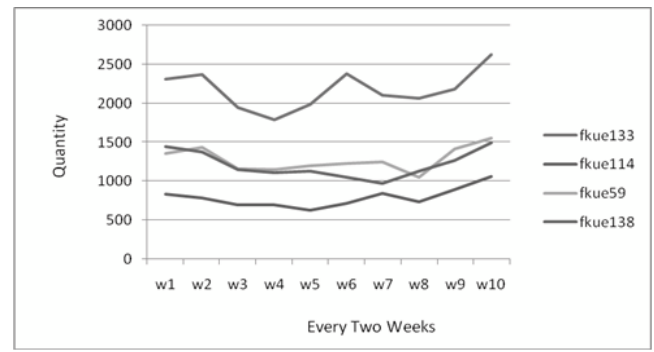

Fig.6 Clustered Chart showing fkue133 has higher transactional volumes compared fkue114 products, fkue59, and fkue138

For future prediction, the very best of 4 selling product items were analyzed. The data set which was used related to the data reduction by considering the way customers buy 4 best product items. The prediction can be seen below:

TABLE I
ANALYSIS THE ACCURACY OF THE PREDICTION FOR 4 VERY BEST PRODUCT SELLING DAILY

| product | p1 | r1 | pr1 | p2 | r2 | pr2 | p3 | r3 | pr3 | p4 | r4 | pr4 | p5 | r5 | pr5 |
|---|---|---|---|---|---|---|---|---|---|---|---|---|---|---|---|
| fkue59 | 14 | 14 | 100% | 12 | 16 | 75% | 11 | 16 | 69% | 16 | 28 | 57% | 17 | 33 | 52% |
| fkue114 | 8 | 8 | 100% | 7 | 4 | 57% | 8 | 12 | 67% | 10 | 9 | 90% | 11 | 12 | 92% |
| fkue133 | 25 | 26 | 96% | 22 | 30 | 73% | 26 | 22 | 85% | 29 | 39 | 74% | 31 | 64 | 48% |
| fkue138 | 14 | 12 | 86% | 14 | 17 | 82% | 16 | 14 | 88% | 16 | 32 | 50% | 22 | 37 | 59% |
| Average | 15.3 | 15 | 96% | 13.8 | 16.8 | 72% | 15.3 | 16 | 77% | 17.8 | 27 | 68% | 20.3 | 36.5 | 63% |

Table one showed proceeds by applying association rules extended to four types of products in six next days and have an accuracy of 100% and accuracy rate degraded

in the next day, though on few next days the accuracy level increase to 100%. The average of accuracy achieve 96%, the degree of accuracy in the last two days is 68% to 63% which is below 70%. Such condition will limit the used extended association rules of the next four days only which is the rules with the accuracy above 70%.

## IV. RECOMMENDATIONS

Further development according this research can be done by:

- FSA-Red algorithms is capable of reducing the high volume of transactional data without having to discard the information related to the analyzed attributes when building the association rules. Rule- making association uses a different algorithm can also use the algorithms FSA-Red in the data reduction process.
- In the research, development Association rules do not consider the value of support due to new products that came later. If support is added as a requirement in making the rules of the association, the new products will be eliminated and the rules of the association regarding the new products will not be created. But in other cases, and different data, the value of support may be considered to obtain a strong association rules with a high confidence level.
- Another prediction algorithm can be examined to extend the reliability of the association rules rather than M5P to get reliable long term prediction in the future.
- The pattern of the time series clustered product groups can be used as another tool to see the rule which can be used for each group of data. Clustering method other than K-means can be used to generate the expected data to cluster better product groups.
- More data is needed which can be use as data testing or data training to get better result for example a year data which expose yearly event or special occasion.

## V. CONCLUSION

In this research, the data preprocessing must be done prior to the high volumes of transactional data before mining the data. The process of transformation will change the composition of the data which making it suitable for mining, while the reduction process will reduce the size of transactional data.

FSA -Red (Feature Selection of Association related to Reduction Process) is used to reduce high volumes of transactional data. Association rules with confidence equal or higher than 70% able to be produced with data that has been reduced using the algorithm FSA-Red. Association rules that have confidence equal or higher than 70% are considered as strong association rules will increase confidence when tested on both the overall the data and next month data.

Furthermore, M5P algorithm can be used to extend the reusability of the association rules which is built by Apriori Algorithm combine with FSA-Red Algorithm for data reduction. The pattern of the time series clustered product groups can be used as another tool to see the rule which can be used for each group of data. Clustering method is reliable to show the truth of extended association rules which been built.

So many association rules with high rate accuracy levels were produced. Business action is selected and made carefully after considering the details of the association rules in time series is patterned on the product group.

### ACKNOWLEDGMENT

This work is partly supported by Widyatama University.